\documentclass[12pt,a4paper]{article}
\usepackage{amsfonts,epsfig}
\usepackage{amsmath,amsfonts}
\usepackage{amssymb,graphicx}
\usepackage{xcolor}
\usepackage{cite}
 \def\pa{\partial}  \def\P{\Phi}
\def\p{\phi} 
\def\k{\kappa} \def\d{\delta}   \def\t{\tilde}
\def\f{\frac}  \def\p{\varphi} \def\k{\kappa}
  \def\t{\tilde} \def\f{\frac} 
 \def\l{\label} \def\e{\varepsilon} 
\def\bt{\beta} \def\la{\lambda}  
\def\m{\mu} \def\n{\nu}   
\def\s{\sigma} \def\S{\Sigma} \def\th{\theta} 
\def\e{\eta}
 \def\be{\begin{equation}}
\def\ee{\end{equation}} \def\ba{\begin{eqnarray}}
\def\ea{\end{eqnarray}}   
\def\s{\sigma} \def\S{\Sigma} \def\th{\theta} 
 \def\be{\begin{equation}}
\def\ee{\end{equation}} \def\ba{\begin{eqnarray}}
\def\ea{\end{eqnarray}}   
\def\s{\sigma} \def\S{\Sigma} \def\th{\theta} 
\def\O{\Omega} 

\def\be{\begin{equation}} \def\ee{\end{equation}}
\def\ba{\begin{eqnarray}} \def\ea{\end{eqnarray}} 
\def\ea{\end{eqnarray}}   
\begin{document}

\vspace{3cm}
\centerline{\large\bf
On the mass function at the inner horizon of a regular black hole }
\vspace{1cm}
\centerline{\bf Mikhail Z. Iofa}
\centerline{Skobeltsyn Institute of Nuclear Physics}
\centerline{Lomonosov Moscow State University}
\centerline{Moscow, 119991, Russia}
\vspace{1cm}

\begin{abstract}
Calculations of the inner mass function of the Hayward regular black
hole with fluxes are reviewed and rederived. We present detailed
calculations of the inner mass function in two forms of the Ori approach (the ingoing flux
is continuous, the outgoing flux is modeled by a thin null shell) and compare
them with calculations in Reissner-Nordstr\"om black hole.
A formal reason of different results is discussed.
The energy density of scalar perturbations propagating from the event
horizon into the Hayward black hole measured by a free falling observer
near the inner horizon is calculated.
\end{abstract}

%%%%%%%%%%%%%%%%%%%%%%%%%%%%%%%%%%%%%%%%%%%%%%%%%%%%%%%%%%%%
\section{Introduction}
%%%%%%%%%%%%%%%%%%%%%%%%%%%%%%%%%%%%%%%%%%%%%%%%%%%%%%%%%%%%%%%

Black hole solutions of general relativity,
Schwarzschild, Reissner-Nordstr\"om (RN), Kerr-Newman, have the central singularity
at $r=0$ which is considered as undesirable in models
of astrophysical black holes.
Regular black holes were proposed as configurations in which the central
singularity is replaced by a non-singular
core \cite{bar,dym,hay,ans,reuter,mod,al}. Regular black holes are static,
spherically-symmetric and satisfy
the weak energy condition.

In this note we discuss the Hayward black hole \cite{hay} which
can be considered as a regularization of the Schwarzschild
solution. Outside the event horizon both geometries have the same
asymptotic form at $r\rightarrow\infty$. The important difference
between Schwarzschild and Hayward black holes is that Hayward
black hole can have no, one double and  two horizons. In the case
of the Hayward black hole with two horizons its causal structure
is similar to that of the RN and Kerr-Newman solutions. In these
black holes the inner horizon  is the  Cauchy horizon, a null
hypersurface beyond which predictability of the theory is lost.

In the process of collapse of a star and formation of a black hole
an outgoing  flux of radiation is produced which,
after a partial reflection at the  potential near the outer
horizon, forms an influx of radiation propagating into black hole
\cite{price}. The influx partially reflects  at the potential in
the vicinity of the inner horizon and produces an outflux. The
inner horizon is a surface of infinite blueshift, and in
\cite{pen,mac1,mac2,gurs1,gurs2, chand, ori-1,ori-2} and many
subsequent papers it was shown that a freely falling observer near
the Cauchy horizon of the RN and Kerr-Newman black holes will see
an unbounded energy density of scalar, electromagnetic and
gravitational fields. This was interpreted as an instability of
the Cauchy horizon
 under external perturbations.

These properties have tended to an expectation that only the
blueshifted ingoing flux (Price radiation tail \cite{price}) near
the inner horizon would result in an increase of the inner mass
function. However, in paper \cite{PI} on the example of the RN
black hole it was shown that only the blueshifted influx is
insufficient for the unbounded increase of the inner mass
function (mass inflation). It was shown that mass inflation appears
only as a combined effect of the incoming and
outgoing fluxes. The mass inflation produces an
increase of curvature at the inner horizon, and instead of the
Cauchy horizon there appears a curvature singularity shielding the
Cauchy horizon. In  paper \cite{PI} the in- and outfluxes were
modeled by the incoming and outgoing charged Vaidia solutions
\cite{vai}, which in turn were modeled by the thin null shells of
lightlike particles crossing without interaction  through each
other. The space-time is divided by the crossing fluxes into four
regions, the metric in each region is characterized by its mass.
Mass inflation of the outgoing flux appears when the ingoing shell
is near the Cauchy horizon. Solution for the inner mass function
was obtained with the use of the Dray-'t Hooft-Redmond (DTR) 
\cite{d-h,red} relation between the masses in the metrics of
four regions. Singularity at the Cauchy horizon was analytically
discussed in \cite{droz,burko,bur-ori,brady} and in a number of
papers cited therein.

In paper \cite{ori}, in the RN black hole with fluxes, the problem
of mass inflation was studied by modeling the outgoing radiation
by a null thin shell, but considering the ingoing Price flux as
continuous (the Ori approach). In this model, starting from  the
Einstein equations and using continuity of the ingoing flux
through the shell, a relation was obtained
connecting the masses of the metrics of spacetimes inside and
outside the shell, which predicted mass inflation
near the inner horizon \cite{ori,and,droz}.

Since the causal structure of the Hayward black hole
is similar to that of the RN black hole, it is natural to discuss,
if, in the Hayward black hole with fluxes, there also appears mass
inflation. This question was discussed within the Ori approach and
using  the generalized DTR construction \cite{BI}, in  \cite{mod}
for the ``loop  black hole'' and in \cite{51,52,bonan} for the
Hayward black hole.  Somewhat surprising result was that, contrary
to the RN black hole, in the case of regular (loop, Hayward) black
holes the Ori approach does not show the mass inflation. However
the use of the (generalized) DTR relation in these models  leads
to mass inflation.

In this work we  review  previous calculations of the inner mass
function and  provide a detailed calculations of the inner mass
function in  two forms of the Ori approach. We compare our results
with calculations in the RN case.

We show that, as in the RN case, in Hayward black hole with an
incoming flux the energy density  near the inner horizon measured by a free falling
observer increases without a bound showing that
both the Hayward and RN black holes are unstable under the
external perturbations.

%%%%%%%%%%%%%%%%%%%%%%%%%%%%%%%%%%%%%%%%%%%%%%%%%%%%%%%%%%%%%%%%
\section{Inner mass function in the Hayward model}
%%%%%%%%%%%%%%%%%%%%%%%%%%%%%%%%%%%%%%%%%%%%%%%%%%%%%%%%%%%%%%%%

The metric of the Hayward black hole is \cite{hay}
\be
\l{1.1}
ds^2 = -f(r) dv^2 +2dvdr +r^2 d\O^2,
\ee
where
\be
\l{1.2}
f(r) = 1 -\f{M(r)}{r} =1- \f{2mr^2 }{2ml^2 +r^3}
.\ee
The metric can have no, one double, or two horizons. We discuss the case
with two horizons.
If the mass $m$ is a function of  retarded or advanced time
$v, m=m(v)$, the metric takes a form
\be
\l{1.2a}
ds^2 =-f(r, v)dv^2 -2  dv dr +r^2 d\O^2
\ee
and there appears an additional component of the energy-momentum tensor
 $T_v^{{}r}$.

In the case there are in- and outfluxes of energy, following the
Ori approach \cite{ori},  the outflux is modeled by a thin null
shell $\S$. The shell divides the interior of the black hole into
an outer $V_+$ and an inner $V_-$ regions. In both parts $V_\pm$
the metric is of the form (\ref{1.2a}) with different $v_\pm$ and
$m_\pm$. The variable $r$ can be introduced
continuous through the shell \cite{BI,P}.

For a metric (\ref{1.2})-(\ref{1.2a}) the Einstein equations take
a form \cite{BI,P} \be \l{1.3} \f{\pa M_\pm}{\pa r}= -4\pi r^2
T^v_v, \qquad \f{\pa M_\pm}{\pa v_\pm }= 4\pi r^2 T^r_v. \ee
Continuity of the metric through  the shell  yields the equation
\be \l{1.4} ({f}_+ dv_+ = {f}_-dv_- )|_{\S} =2dr. \ee Below we
write $v_+\equiv v,\,\, f_+ (v_+)=f (v)$.

Continuity of the flux across the shell $[T_{\m\n}n^\m
n^\n]=0,\,\, [A] =A_+ -A_- , \,\, n^\m $ being a
normal to the shell, is expressed as \be \l{1.5} \f{T_{v_+
v_+}}{{f}^2_+}=\f{T_{v_- v_-}}{{f}^2_-}\bigg |_\S .\ee From
Eq.(\ref{1.4}) variable $v_-$ is determined  as a function of $v$
\be \l{1.6} \f{d{v}_-(v)}{dv}=\f{ f(v)}{\t{f}_-(v)}\bigg|_\S ,\ee
where $\t{f}_-(v)=f_-(v_-(v))$. In the same way,  $\t{m}_-(v)=
m_-(v_-(v))$. We have \be \l{1.7} \f{\pa \t{M}_-(v)}{\pa v}=\f{\pa
M_-(v_-)}{\pa v_-}\f{{f}(v)}{\t{f}_-(v)} \bigg|_\S .\ee From
Eqs.(\ref{1.5}) and (\ref{1.7}) we obtain \be \l{1.71}
\f{1}{f(v)}\f{\pa M_+}{\pa v}\bigg|_\S = \f{1}{\t{f}_-(v)}\f{\pa
\t{M}_-}{\pa v}\bigg|_\S .\ee Eq.(\ref{1.71}) will be used to
obtain the mass function $\t{m}_-(v)$.

%%%%%%%%%%%%%%%%%%%%%%%%%%%%%%%%%%%%%%%%%%%%%%%%%
\subsection{ Mass function from continuity of flux across the shell}
%%%%%%%%%%%%%%%%%%%%%%%%%%%%%%%%%%%%%%%%%

In the case of a black hole with the Price influx \cite{price}, the mass function
in the region $V_+$  is
$m_+ =m_0 -\d m_{pr}$, where $m_0$ is the mass without the Price flux,
and $\d m_{pr}(v)=\bt/v^p$, where $p\geq 12$.
We assume that $m_0\gg l$.

Without the Price flux, the  horizons of the black
hole are determined from the equation $f(r, m_0 )=0$, or,
equivalently, from the equation
 \be
\l{2.1}
r^3 -2m_0 (r^2 -l^2)=0
.\ee
 The outer horizon is located at
${r}_+\simeq 2m_0 - l^2/2m_0 +\cdots $, the inner horizon is at
\be \l{2.1a} {r}_- \simeq
l\left(1+\f{l}{4m_0}+\f{5}{2}\left(\f{l}{4m_0}\right)^2+\cdots\right)
\equiv l(1+\e +5\e^2/2+\cdots) .\ee In the case  with fluxes,
the locations of the  horizons are \be \l{2.1b}
r_+(v)\simeq {r}_+ -\d r_+(v),\qquad r_-(v)\simeq {r}_- +\d
r_-(v). \ee
 In the first order in $\d m_{pr}$, $ \d r_-$ is determined from the equation
$$
 f_{,r}({r}_-, m_0)\d r_- -f_{,m}({r}_-, m_0)\d m_{pr}=0,
%\d r_- =\f{2\d m_{pr}(\t{r}_-^2 -l^2)}{4m_0\t{r}_- -3\t{r}^2}.
$$
which gives
\be
\l{2.1c}
\d r_-= \d m_{pr}\,f_{,m}({r}_-, m_0)/f_{,r}({r}_-, m_0).
\ee
The shell modeling the outflux is located at
the radius $r_{shell} =r_s$, and  in
 vicinity of the inner horizon we can write
 $r_s ={r}_- +y(v),\,\,  \t{r}_- > y(v)$.
The location of the shell  is
 determined  by the null geodesic equation (\ref{1.4})
\be \l{2.2} 2\dot{r}_s (v) =f(r_s, m_+(v))=f({r}_- +y(v), m_0 -\d
m_{pr}(v)). \ee
The dot denotes the derivative with
respect to the variable $v$. In  the first order in $y$ and $\d
m_{pr}$, we obtain \be \l{2.4} 2\dot{y} = f_{,r}({r}_- ,m_0) y -
f_{,m}({r}_- ,m_0)\d m_{pr} ,\ee where \ba \l{2.5} f_{,r}(m_0
,r_-)=2m_0 {r}_- \f{ {r}_-^3 -4m_0 l^2 } { ({r}_-^3 + 2m_0 l^2 )^2
}
\simeq -\f{2}{l}( 1-4\e ), \\
\l{2.6} f_{,m}(m_0 ,r_-)= -\f{2{r}_-^5}{({r}_-^3 +2m_0
l^2)^2}\simeq - \f{l}{2m_0^2 }(1+\e ). \ea Solving (\ref{2.4}), we
have \be \l{2.8} y(v)= e^{-v|f_{,r}|/2}\left(C - \int^v\,dv
e^{v|f_{,r}|/2} \d {m}_{pr}\f{f_{,m}}{2} \right). \ee In the limit
$v\rightarrow\infty$, $y(v)$ approximately is \be \l{2.10}
y(v)\simeq \f{l^2}{4m_0^2}\d m_{pr}(1+5\e)-\d
\dot{m}_{pr}\f{l^3}{4m_0^2}(1+9\e) . \ee
If the location of the shell is determined with respect to
 the $v$-dependent horizon $r_-(v)$, $r_s =r_-(v) +z(v)$, $z(v)$ is determined
from the geodesic equation \be \l{2.12}
2(\dot{z}+\d\dot{r}_-)=f_{,r}(r_-,m_0)(\d r_-
+z)-f_{,m}(r_-,m_0)\d m_{pr} ,\ee where $\d r =\d
m_{pr}(l^2/4m_0^2) (1+5\e)$. Asymptotic solution of (\ref{2.12})
is \be \l{2.13} z(v) =e^{-v|f_{,r}|/2}\left(C - \int^v\,dv
e^{v|f_{,r}|/2} \d m_{pr}\f{l^2}{4m_0^2} (1+5\e)\right) \simeq
-\d\dot{m}_{pr} \f{l^3}{4m_0^2}\left(1+9\e\right) .\ee The terms
proportional to $\d m_{pl}$ have canceled. The
location of the shell in both calculations is \be \l{2.11} r_s
(v) \simeq {r}_- +\f{l^2}{4m_0^2}\d{m}_{pr}(1+5\e
)-\f{l^3}{4m_0^2}\d\dot{m}\left(1+9\e\right) .\ee

 Using (\ref{1.71}), let us find the inner mass function $\t{m}_- (v)$.
First, we  calculate
\ba
\l{2.14}
R_+ \equiv \f{1}{f(v)}\f{\pa M_+}{\pa v}=\f{r_s^6 \dot{m}_+ }{(r_s^3 +2l^2 m_+ )
(r_s^3 -2m_+ (r_s^2 -l^2 ))}=
\f{r_s^6 \dot{m}_+ }{2\dot{r_s}(2m_+ l^2+ r_s^3)^2}.
\ea
Here we have used the geodesic equation of the shell written as
$r_s^3 -2m_+ (r_s^2 -l^2 )=2\dot{r_s}(2m_+ l^2+ r_s^3).$
Using (\ref{2.11}) and retaining in (\ref{2.14})
 the leading in $\d m_{pr}$ terms , we have
\be
\l{2.15}
R_+ \simeq   -\f{r_s^6}{2 l^6(1+ 9\e )}.
\ee
The equation for the inner mass function $\t{m}_-$ is
\be
\l{2.16}
R_-\equiv \f{1}{\t{f}_-(v)}\f{\pa \t{M}_-}{\pa v} =
\f{ r_s^6 \dot{\t{m}}_- }{-4\t{m}_-^2 l^2 (r_s^2 -l^2 )+ 2\t{m}_-
 r_s^3 (2l^2 -r_s^2 ) +r_s^6} =-\f{r_s^6}{2 l^6 (1+9\e )}.
\ee
Using the relations
\be
\l{2.16a}
r_s^2 -l^2 \simeq l^2 [(1+\e +5\e^2 /2+\cdots)^2 -1]\simeq
\f{l^3}{2m_0}(1+3\e ),\qquad 2l^2 -r_s^2\simeq l^2(1-2\e)
,\ee
we rewrite Eq.(\ref{2.16}) as
\be
\l{2.17}
\dot{\t{m}}_- = \f{1}{1+9\e}\left[\f{\t{m}_-^2}{2m_0 l}(1+3\e)-\f{\t{m}_-}{l}
(1+\e )-\f{1}{2}(1+6\e )\right]
,\ee
and further is manipulated to
\ba
\l{2.18}
&{}&\dot{\t{m}}_-=
\f{1-6\e}{m_0 l}\left[\left(\t{m}_- -\f{m_0}{2} (1-2\e )\right)^2
 -\f{m_0^2}{4} (1-2\e )^2- \f{m_0 l}{2}(1+3\e) \right]\simeq \\\nonumber
&{}&\simeq \f{1-6\e}{m_0 l}\left[\left(\t{m}_- -\f{m_0 }{2} (1-2\e )\right)^2
 -\left(\f{ m_0 }{2} (1+2\e )\right)^2\right]
.\ea
Integrating Eq.(\ref{2.18}), we have
\be
\l{2.19}
\f{1}{m_0 (1+2\e )}\ln
\bigg|\f{(\t{m}_- -m_0  (1-2\e )/2)- m_0 (1+2\e )/2}
{(\t{m}_- -m_0  (1-2\e )/2)+ m_0 (1+2\e )/2 }\bigg|= C {v/2l}
.\ee
In the limit $v\rightarrow \infty$ we obtain
\be
\l{2.20}
\t{m}_- = -2m_0 \e =-\f{l}{2}
.\ee

The negative value for $m_-$ was discussed in \cite{mod} for the loop black
hole and in \cite{52} for the Hayward black hole, and in particular it was  noted
 that the inner mass function $m_- (v)$
is not a directly measurable quantity and the result may be an artifact of
parametrization.

%%%%%%%%%%%%%%%%%%%%%%%%%%%%%%%%%%%%%%%%%%%%%%%%%%%%%%%%%%%%%%
\subsection{Mass function in double null coordinates}
%%%%%%%%%%%%%%%%%%%%%%%%%%%%%%%%%%%%%%%%%%%%%%%%%%%%%%%%%%%%%

Let us calculate the inner mass function following the original
Ori approach \cite{ori}. As above, the shell divides the interior
of the black hole into two regions $V_\pm$ with different $v_\pm$
and $m_\pm$; $r$ is continuous across the shell. In double null
coordinates the metric is $ds^2 =-2e^{\s}dUdV +r^2 d\O$.
The coordinate $U$ is set equal to zero at the shell.
Since the shell is pressureless, it is possible to introduce an
affine parameter $\la$ at both sides of the shell \cite{BI,P}.
Coordinate $V$  is equal to $\la$ along the shell. Location of the
shell $ r_s = r_- +y(v) $ (see(\ref{2.11})), is written  as $r_s
=R(\la)=r(V=\la, U=0 )$.
 It is supposed that the shell reaches the inner horizon
at $v\rightarrow\infty$, or, equivalently, at  $\la =0$.
In the following we write  $v_+=v$.

The geodesic equation of the shell is \be \l{4.1} R'/v'_\pm =
\f{1}{2}f_\pm (m_\pm (v), R), \ee where the prime
denotes the derivative with respect to $\la$. One introduces \be
\l{4.2} z_\pm =R/v'_\pm .\ee Differentiating  Eq.(\ref{4.2}),
$z'_\pm = R'/v'_\pm -Rv_\pm''/{v'_\pm}^2$,
 and using the geodesic equation for $v(\la )$
\be
\l{4.3}
v''_\pm+\f{1}{2}f_{\pm,r}{v'_\pm}^2 =0
,\ee
we obtain an equation
\be
\l{4.4}
2 z'_\pm = f_\pm +R f_{\pm , r}
.\ee
At the $(+)$ side of the shell  Eq.(\ref{4.4}) takes a form
\be
\l{4.5}
z'_+ =\f{1}{2}\left(1- 12\f{m_+^2 R^2 l^2}{(R^3 +2m_+  l^2)^2}\right),
\ee
or
\be
\l{4.5a}
z'_+ \simeq -1 -2\left(l/4m_0 +\d m_{pr} /m_0 + \d y/l \right)
.\ee
Here we have set  $R(\la )= r_- +y(R,v) $ and
$m_+ =m_0 - \d m_{pr}$.
Integrating  (\ref{4.5}), we obtain $z_+ \simeq Z_+ -\la$.
From (\ref{4.2}) it follows that
\be
\l{4.6}
v_+ =\int^\la d\la \f{R}{z_+} \simeq r_- \ln\f{1}{\la}.
\ee
In (\ref{4.6}) $Z_+$ was set to zero to have $v_+\rightarrow\infty$ for $\la
\rightarrow 0$.

Differentiating (\ref{4.1}) with respect to $\la$, we have
$ v_{\pm}''=2(R''f_{\pm} -R'f_{\pm}')/f^2_{\pm}$.
Substituting
$v''$ in (\ref{4.3}),  we obtain an equation for $f$ (See also
\cite{52})
\be
\l{4.7}
f(R)\,\f{R''}{R'}=f'(R) -f_{,r}(R)R'
.\ee
In the case $f=f_-$,
$$
f_- (\t{m}_-(v),R) =1-\f{2\t{m}_-(v) R^2}{R^3 +2\t{m}_-(v)l^2},
$$
transforming  (\ref{4.7}), we obtain
\be
\l{4.7a}
\f{R''}{2R'}f_-(\t{m}_-(v), R) =-\f{\t{m}'_- R^5}{(R^3 +2\t{m}_-  l^2)^2}
.\ee
Substituting $R(v(\la )) =r_s \simeq r_- + \d m (v) l^2/4m_0^2$, we have
$$
\f{R''}{R'}= \f{ p(p+1)v^{-p-2}{v'}^2 -pv^{-p-1}v'' }
{ - p v^{-p-1}v'}=-(p+1)\f{v'}{v}+ \f{v''}{v'} \simeq -\f{1}{\la}
.$$
Eq.(\ref{4.7a}) takes a form
\be
\l{4.8}
\t{m}'_-  = -\f{1}{2\la\, R^5} [4\t{m}_-^2 l^2 (R^2 -l^2 )-2\t{m}_-R^3
(2l^2 -R^2 ) -R^6 ].
\ee
Noting that
$$ \t{m}'_- =\f{d\t{m}(v)}{dv}\left(-\f{r_-}{\la}\right), $$
 we have
\be
\l{4.9}
\dot{\t{m}}_- =\f{2l^2 (r_-^2-l^2 )}{r_-^6}\left[{\t{m}}_-^2
-2{\t{m}}_- \f{r_-^3 (2l^2 -r_-^2)}{4 l^2 (r_-^2-l^2 )}-
\f{r_-^6}{4 l^2 (r_-^2-l^2 )}\right].
\ee
In Eq.(\ref{4.9}) we recognize the structure of Eq.(\ref{2.16}).
With the use of (\ref{2.16a}), Eq.(\ref{4.9}) takes the same form as
(\ref{2.18}).

%%%%%%%%%%%%%%%%%%%%%%%%%%%%%%%%%%%%%%%%%%%%%%%%%%%%%%%%%%%%%%%%
\section{Hayward versus RN black hole}
%%%%%%%%%%%%%%%%%%%%%%%%%%%%%%%%%%%%%%%%%%%%%%%%%%%%%%%%%%

Let us compare  the Hayward black hole with
 the  RN black hole with the Price flux.

The function $f(v)$ in (\ref{1.1}) is \be \l{3.1}
f(v)=1-\f{2m(v)}{r}+\f{e^2}{r^2} .\ee We assume that $m(v)\gg e^2
$. The mass of the black hole, $m_+$, is $m_0 -\d m_{pr},\,\, m_0
\gg \d m_{pr}$. Without the Price flux, the locations
of the horizons are determined from the equation $f(r, m_0 )=0$.
The outer and inner horizons are located at $\t{r}_+ =m_0
+\sqrt{m_0^2 -e^2}$ and $\t{r}_- =m_0 -\sqrt{m_0^2 -e^2}$. In
the case with flux, locations of horizons are \be \l{3.01}
r_+=\t{r}_+\left(1 -\f{ \d m_{pr} }{\sqrt{m_0^2
-e^2}}\right),\qquad r_-=\t{r}_-\left(1 +\f{ \d m_{pr}
}{\sqrt{m_0^2 -e^2}}\right) .\ee The  shell modeling the outflux
is located in the vicinity of the inner horizon at $r_s (v)={r}_-
(v) +y(v)$. Below we use the notations of the preceding section.

The location of the  null shell is determined by the
geodesic equation $2\dot{r_s}= f(r_s, m_+ )$. Noting that
$f(\t{r}_-, m_0)=0$ and expanding the function $f(r_s, m(v))$ to
the first order in $y$ and $\d m_{pr}$, we obtain \be \l{3.2}
2\left(\dot{y}+\f{\d \dot{m}_{pr} }{\k\t{r}_-}\right) =
f_{,r}(\t{r}_- ,m_0 )\left(y +\f{ \d m_{pr} }{\k\t{r}_-}\right)-
f_{,m}(\t{r}_- ,m_0 )\d m_{pr}, \ee where \be \l{3.21} r_s
=\t{r}_- +y +\f{\d m_{pr}}{k\t{r}_- }; \qquad \k =\f{\sqrt{m_0^2
-e^2}}{\t{r}_-^2};\qquad f_{,r}(\t{r}_- ,m_0 )=-2\k ; \qquad
f_{,m}(\t{r}_- ,m_0 )=-\f{2}{\t{r}_-} .\ee Substituting
expressions (\ref{3.21}) into (\ref{3.2}), we rewrite
Eq.(\ref{3.2}) in  a form \be \l{3.3} \dot{y} +\k y \simeq
-\f{\d\dot{m}_{pr}}{\k\t{r}_-} .\ee Note that as in (\ref{2.11})
the terms proportional to $\d m_{pr}$ have canceled. Solution of
Eq.(\ref{3.3}) is \be \l{3.4} y(v)=e^{-\k v}\left(C -\int^v dv
e^{+\k v}\f{\d\dot{m}_{pr}}{\k\t{r}_-} \right). \ee
%Note that (\ref{3.4}) is similar to (\ref{2.6}).
In the limit $v\rightarrow\infty$ solution (\ref{3.4}) is simplified to
$$
y(v)\simeq -\f{ \d\dot{m}_{pr} }{\t{r}_-\k^2}.
$$
The null shell is located at \be \l{3.5} r_s = \t{r}_- +\f{\d
m_{pr}}{\t{r}_- \k}-\f{\d\dot{m}_{pr}}{\t{r}_-\k^2} .\ee Let us
find the inner mass function. With the use of
relations (\ref{1.3})-(\ref{1.5}) continuity of the flux across
the shell is written as \be \l{3.6} \f{\dot{ \t{m} }_- (v_- (v))
}{ \t{f}_- ( \t{m}_-,v )}=\f{\dot{m}_+(v)}{f(v)}. \ee Using the
geodesic equation to determine the location of the
shell, $f(v)=2\dot{r}_s$, and writing $f_-(\t{m}_- ,v)=f(v)+2(m_+
-\t{m}_-)/r_s$, we obtain Eq.(\ref{3.6}) in a form \be \l{3.7}
\f{\dot{\t{m}}_-(v)}{2\dot{r}_s + 2(m_+ - \t{m}_- )/r_s }=
-\f{\d\dot{m}_{pr}}{2\dot{r}_s}, \ee From (\ref{3.5}), we have \be
\l{3.8}
\f{\d\dot{m}_{pr}}{2\dot{r}_s}=\f{\k\t{r}_-}{2}\left(1+\f{p+1}{v\k}
\right) .\ee Neglecting in the l.h.s. of (\ref{3.7}) the small
term $\dot{r}_s$, we obtain \be \l{3.9} \dot{\t{m}}_- \simeq (
\t{m}_- -m_0 )\k \left(1+\f{p+1}{v\k} \right). \ee Here we meet
the crucial difference from the Hayward black hole: Eq.(\ref{3.9})
is of the first order in $\t{m}_-$ while (\ref{2.18}) contains
$\t{m}_-$ quadratically. Solving (\ref{3.8}), we obtain \be
\l{3.10} \t{m}_-(v)= e^{\k v+(p+1)\ln v }\left(C-\k\,m_0 \int^v dv
\left(1+\f{p+1}{v\k} \right)e^{-\k v-(p+1)\ln v}\right) \simeq C
e^{\k v}v^{(p+1)}- m_0 .\ee

%%%%%%%%%%%%%%%%%%%%%%%%%%%%%%%%%%%%%%%%%%%%%%%%%%%
To finish the comparison of calculations of the inner
mass in the RN and Hayward  black holes, we calculate  the inner
mass in the RN black hole in the Ori approach as in Sect.2.2.

Eq.(\ref{4.4}) for the (+) side,
 $2 z'_+ = f_+ +R f_{+ , r}$, gives
\be
\l{3.101}
2z'_+ =1-\f{e^2}{r_s^2}\simeq 1-\f{e^2}{\t{r}_-^2}=-2\k \t{r}_-
,\ee
(for definitions of $\t{r}_-$ and $\k$ see (\ref{3.01}) and (\ref{3.21}))
and we obtain
\be
\l{3.102}
v_+ \equiv v =\f{1}{\k}\ln\f{1}{\la}.
\ee
Eq.(\ref{4.7}) for the case of the RN black hole has a form
\be
\l{3.11}
-\f{2\t{m}_-'(v(\la))}{R} =\f{R''}{R'}f(R, \t{m}_- (v) )\big|_{R=r_s}
.\ee
Substituting in (\ref{3.11}) the relations
$$
-\f{2\t{m}_-'}{r_s}\simeq
 \f{2\dot{\t{m}}_-}{\t{r}_- \k \la},\qquad \f{R''}{R'}=\f{r_s''}{r_s'}=
-\f{1}{\la} \left(1+\f{p+1}
{\ln \la}\right)
$$
and
$$
f(r_s, \t{m}_- ) =
\left(2\dot{r}_s + \f{2 (m_+ -\t{m}_- )}{r_s} \right)
,$$
we obtain the equation
\be
\l{3.12}
\dot{\t{m}}_- \simeq ( \t{m}_- -m_0 )\k \left(1+\f{p+1}{v\k} \right),
\ee
which coincides with (\ref{3.9}).

%%%%%%%%%%%%%%%%%%%%%%%%%%%%%%%%%%%%%%%%%%%%%%%%%%%%%%%%%%%%%%%%%%%
\section{Instability of the inner horizon under external perturbations}
%%%%%%%%%%%%%%%%%%%%%%%%%%%%%%%%%%%%%%%%%%%%%%%%%%%%%%%%%%%%%%%

In this section we consider propagation of scalar field  in
a neighborhood of the Cauchy horizon and show that the power-law tails
entering the black hole as seen by a free falling observer diverge at the inner horizon.

The problem of external perturbations for the Hayward black hole
 is discussed in a similar way as
in the case of the RN black hole \cite{sand,gurs1,gurs2,chand,ori-1},
because the causal structures of both metrics are similar.

To set the problem, we consider the Hayward metric
$$
ds^2 =-f(r)dt^2 +\f{dr^2}{f(r)} +r^2 d\O^2
,$$
where
\be
\l{5.6}
f(r)=f_{RN}\,g(r)=-\f{(r_+ -r)(r-r_-)}{r^2}\f{r^2 (r -\t{r})}{r^3+2ml^2}
.\ee
Here $\t{r} =-r_-( -m)$ is the negative root of the equation
$$
r^3 -2m(r^2 -l^2)= 0 =[(-r)^3 -(-2m)(r^2 -l^2)].
$$
$g(r)$ is a bounded function without  zeroes   and poles at $r>0$.

In the region $(r_+,r_-)$ one introduces  the tortoise variable
$r_* =-\int dr/f(r)$
\ba
\nonumber
%&{}&
r_* = \int\,dr \f{r^3+2ml^2}{(r_+ -r)(r-r_-)(r-\t{r})}=\hspace*{10cm}\\\l{5.2}
%&{}&
-r -A_1 (r_+^3 +2ml^2)\ln (r_+ -r) -A_2 (r_-^3 +2ml^2)\ln (r -r_-)
-A_3(\t{r}^3 +2ml^2)\ln(r -\t{r}) +const
,\ea
where
\be
\l{5.21}
A_1= \f{1}{(r_+ -r_-)(r_+ -\t{r})}, \qquad
A_2= \f{1}{(r_+ -r_-)(\t{r}-r_-)} , \qquad
A_3= \f{1}{(r_+ -\t{r})(r_- -\t{r})}
.\ee
Assuming as in Sect.2 that $m\gg l$, so that $r_+ \simeq 2m,\,r_-\simeq l,\,
\t{r}\simeq -l $, and
$$
 A_1\simeq 1/(2m)^2,\,\,A_2\simeq -1/4ml ,\,\,A_3\simeq 1/4ml ,
$$
 we obtain
\be
\l{5.22}
r_* \simeq -r -2m\ln (r_+ -r) +\f{l}{2}\,\ln (r -r_-)-\f{l}{2}\,\ln (r -\t{r}).
\ee
In the limit $r\rightarrow r_-$ we have
\be
\l{5.22a}
r_* \simeq \f{l}{2}\ln\f{r-r_-}{l}.
\ee
Defining the null coordinates $v= -r_* +t,\,\,u=-r_* -t$, the left and right
branches of the Cauchy horizon are
 the hypersurfaces $ (r_- ,u=\infty)$ and $(r_- , v=\infty)$.
Propagation of the scalar  field $\Phi (x)$ is described by the
wave equation $\Phi_{;\m;\n}f^{\m\n} =0$, where $f_{\mu\n}$ are
components of the metric (\ref{5.6}). To solve the equation, the
field $\Phi (x)$ is expanded in spherical harmonics \be \l{5.31}
\Phi (x) =\int e^{-ikt}Y_{lm}(\th,\p
)H_{lm}(k)\f{\p_{klm}(r)}{r}dk .\ee The functions $\p (r_{_*}(r))$
(below the indices $klm$ are omitted) satisfy the equation \be
\l{5.4} \f{d^2 \p (r_* )}{d r^2_{_*}}+[ k^2 -V_l (r_{_*}(r))]\p
(r_*) =0, \ee where the potential $V_l$ is
$$
V_l (r_{_*}(r))=- f(r)\left[\f{l(l+1)}{r^2} +\f{1}{r}\f{df(r)}{dr} \right].
$$
The function $H(k)$ in (\ref{5.31}) is determined by the initial
data $h(v)$  given at the branch $(r_+ ,u= -\infty )$ of the outer
horizon \be \l{5.13} H(k) =\f{1}{2\pi}\int\,e^{ikv} h(v) dv .\ee
The solutions of (\ref{5.4}) $\p (r_*)$ which  have
the asymptotic form  $e^{-ikt}\p (r_*)\sim e^{-ikv}$ at the $r_+$
horizon, at the $r_-$ horizon are
$$
e^{-ikt}\p (r_*)\sim A(k)e^{-ikv} +B(k)e^{iku}\qquad r_*\rightarrow -\infty,
$$
where $A_{lm}^2 -B_{lm}^2 =1$.

The ingoing field $\P(r_*,t)$ propagates inside the black hole and
near $r_-$ is scattered in fluxes $X(v)$ and $Y(u)$
\be
\l{5.10}
 \P(r_*,t)  \rightarrow X(v)+Y(u)=
\int dk H(k)( A(k)-1)e^{-ikv}+\int dk H(k) B(k)e^{iku}. \ee In the
limit $v,u\rightarrow\infty$, the main contribution to $X(v)$ and
$Y(u)$ comes from integration in a vicinity of $k=0$
\cite{gurs1,gurs2}. For the Price power-law tail  $h(v)=\d m_{pr}=
\bt \th (v-v_0 )v^{-p}$ one obtains \ba \nonumber
&{}& X(v)=\bt v^{-p} (A(0) -1),\qquad  v\rightarrow\infty,\\
&{}& Y(u)=\bt u^{-p} B(0), \qquad  u\rightarrow\infty.
\ea
The fields $X(v)$ and $Y(u)$ are finite on the Cauchy horizon.

Let us find what energy density  measures a free falling observer
in a vicinity of the inner horizon. The velocity components of the radially
falling observer are
\cite{chand,sand}
\be
\l{5.3}
U^t = \f{E}{f(r)}, \qquad U^{r} =-\sqrt{E^2 - f(r)},
\ee
and $U^{r_*}=\sqrt{E^2 - f(r)}/f(r)$.
The flux seen by a free falling observer is
\ba
\nonumber
&{}&
{\cal F}= U^i \P_i =U^t \Phi_{,t}+U^{,\, r_*}\Phi_{r_*}=
\f{E}{f(r)}(X_{,v}-Y_{,u}) +\f{\sqrt{E^2 - f(r)}}{f(r)}(-X_{,v}-Y_{,u})=\\
&{}&\f{X_{,v}}{f(r)}(E-\sqrt{E^2 - f(r)})-\f{Y_{,u}}{f(r)}(E+\sqrt{E^2 - f(r)})
.\ea
In the limit $r\rightarrow r_-$, or $r_*\rightarrow -\infty$,
 the function $f(r)$ vanishes:
$$f(r)\sim 2\f{r-r_- }{l} \sim \f{e^{2 r_*/l} }{2}.$$
If $E>0$, the flux is
\be
\l{5.15}
 {\cal F}\simeq -\f{f(r)}{2E} X_{,v}-\f{2E}{f(r)}Y_{,u}
.\ee
At the branch $(r_- ,u\rightarrow \infty )$ the first term is finite, and the
second increases.
If $E<0$, we obtain
\be
\l{5.14}
{\cal F}\simeq  \f{2|E|}{f(r)} X_{,v} +  \f{f(r)}{2|E|}Y_{,u}
.\ee
At the branch $(r_- , v\rightarrow \infty )$ the first term increases,
and the second is finite.
The fluxes measured by a free falling observer are
\ba
\l{5.11}
&{}& E>0:\qquad {\cal F}|_{(r_- ,u\rightarrow\infty)}
=-2E \bt p u^{-p-1} B(0)e^{2u/l},\\
\l{5.12}
&{}& E<0:\qquad {\cal F}|_{(r_- ,v\rightarrow\infty)}
=2E \bt p v^{-p-1} (A(0)-1)e^{2v/l}.
\ea
It is seen that the observed fluxes exponentially diverge at the inner horizon.

%%%%%%%%%%%%%%%%%%%%%%%%%%%%%%%%%%%%%%%%%%%%%%%
\section{Conclusions and discussion}
%%%%%%%%%%%%%%%%%%%%%%%%%%%%%%%%%%%%%%%%%%%%%%%%%

In this work we studied the inner mass function in the Hayward
model of regular black hole with fluxes and compared it with the
RN black hole. We assumed that the mass of the black hole without
fluxes, $m_0$, is much larger than the core parameter $l$.
Assuming the validity of the classical treatment, we
consider the core parameter $l$ larger than the Planck length
$l_p$.

We calculated the inner mass via  two methods, the first based on
continuity of the flux across the shell and the second one using
the original Ori approach \cite{ori} (to be precise, both methods
are  within the Ori approach, because in both methods the incoming
flux was taken as the continuous Price flux, and the outgoing flux
was modeled by a pressureless null shell). Both methods give a
finite negative value for the inner mass function. The inner mass
is not a directly measurable quantity, and in \cite{mod,52} it was
suggested that this result is an artifact of the
parameterization. However, a good parameterization is not known.

Formally, different behavior of the inner mass functions in RN and Hayward
black holes is traced to a following.
Schematically, in the RN case, the equation for the inner mass is
$$
\f{dm}{dv} -cm =-\d m_{pr}
,$$
where $c>0$, leading to the exponential grows in $v$. For the Hayward
black hole, we have
$$
\f{dm}{dv} = (m-c)^2  -a^2, \qquad c,a>0,
$$
which gives
$$
\bigg|\f{(m-c)-a}{(m-c)+a}\bigg|=Ce^{2av}
.$$
In the limit $v\rightarrow\infty$ the solution for $m(v)$ is
$m=c-a$ which for the specific values of $c$ and $a$ gives $m<0$.

Another approach used in \cite{PI,mod} to find the inner mass
function, is  based on the Dray-'t Hooft-Redmond (DTR)
\cite{d-h,red} relation. In this approach, the  fluxes in the
interior of the black hole are modeled as thin shells. The DTR
formula provides a relation  between the masses of the metrics of
the ingoing and outgoing spherical shells in the regions between
the shells before and after collision. In \cite{PI}, in the case
of the RN black hole, the DTR relation was derived from the system
of the Einstein equations. For the loop and Hayward black holes
one must use the generalized DTR formula \cite{BI,mod} which does
not require the use of the Einstein equations.
    It appears that in all cases of  RN, Hayward and loop black holes
the generalized DTR relation shows the divergence of the mass function
of the spacetime near the Cauchy horizon
after the shells have crossed.  It was noted that DTR relation
being  nonperturbative accounts for nonlocal and nonlinear effects \cite{BI}.

However, the DTR approach is  not directly comparable with the Ori
approach and there lacks a clear connection between the DTR
formula and the Ori-like approaches.

In \cite{ham} it was shown that in a system with crossing streams
inside a black hole generated by accretion appears mass inflation.
The streams propagate in the background of spherically-symmetric
space-time and mass inflation appears as the 4-velocities of the
streams increase at the approach to the inner horizon. Because of
the Lorentz boost of  4-velocity of the observed flux with respect
to the observing flux, the counter-streaming velocity of fluxes
exponentiates along with the center-of-mass energy density of the
streams, causing  the increase of the interior mass. However, in
paper \cite{ham} calculations were made with the specially
constructed   metric, and a concrete reformulation of  the results
of \cite{ham} to models with the metrics of the form (\ref{1.1})
is a problem.

Because of the similar causal structure of the metrics of the Hayward and RN
black holes, in both cases propagation of external perturbations is also similar.
If perturbation is the Price flux, in both cases a free falling observer approaching the inner horizon
measures an increasing energy flux. This property is interpreted as an instability
of the inner horizon. However, this effect is not directly connected
with the inner mass inflation.

\vspace{3mm}
{\large\bf Acknowledgments}

I thank M. Smolyakov and I. Volobuev for discussion and valuable comments.

The research was carried out within the framework of the scientific program
of the National Center for Physics and Mathematics,
 the project ``Particle Physics and Cosmology'', and was partially supported
by the Project 01201255504 of the Ministry of Science
and Higher Education of Russian Federation.

\end{document}